\documentclass[12pt]{iopart}
\usepackage{graphicx}
\usepackage{color}

\newcommand{\eqref}[1]{(\ref{#1})}

\begin{document}

\title[Metastable Clusters and Channels Formed by Active Particles]{Metastable Clusters and Channels Formed by Active Particles with Aligning Interactions}
\author{Simon Nilsson$^{1,2}$, Giovanni Volpe$^2$}

\address{$^1$ Department of Physics, Chalmers University of Technology, G\"oteborg, Sweden}
\address{$^2$ Department of Physics, University of Gothenburg, G\"oteborg, Sweden}

\begin{abstract}
We introduce a novel model for active particles with short-range aligning interactions and study their behaviour in crowded environments using numerical simulations. When only active particles are present, we observe a transition from a gaseous state to the emergence of metastable clusters as the level of orientational noise is reduced. When also passive particle are present, we observe the emergence of a network of metastable channels.
\end{abstract}

\section{Introduction}

Collective behaviours often emerge in systems constituted by individuals capable of motion and of interaction with their environment (active particles) \cite{bechinger2016active}. While the ensuing complex behaviours seem to pose a daunting task for our understanding, their defining characteristics can be captured by simple models, where complex collective behaviours emerge even if each active particle follows very simple rules, senses only its immediate surroundings, and directly interacts only with nearby particles, without having any knowledge of an overall plan. In particular, it has been found that systems of interacting active particles give rise to robust and universal emergent behaviours occuring at many different length and time scales with classical examples going from swarms of bacteria and syntethic microswimmers, to schools of fish, flocks of birds and human crowds \cite{vicsek2012collective,gautrais2012deciphering}.

The first model for collective motion was introduced to model the swarm behaviour of animals at the macroscale. In 1987, Reynolds introduced the Boids model to simulate the aggregate motion of flocks of birds, herds of land animals, or schools of fish within computer graphics applications \cite{reynolds1987flocks}. Then, in 1995, Vicsek and co-authors introduced the Vicsek model as a special case \cite{vicsek1995novel}, where a swarm is modeled by a collection of active particles moving with constant speed and tending to align with the average direction of motion of the particles in their local neighborhood. Later, several additional models have been introduced to capture the properties of collective behaviours \cite{chate2008modeling, barberis2016large, mijalkov2016engineering, matsui2017noise, cambui2017finite}. 

Several systems featuring complex collective behaviours have also been realised experimentally. Motile bacteria have been shown to form vortices and other spatial patterns \cite{czirok1996formation}. Artificial active particles have been shown to cluster and form crystal-like structures \cite{palacci2013living, theurkauff2012dynamic, ginot2015nonequilibrium}.

Beyond its intrinsic scientific interest, a deep understanding of collective behaviours can contribute to applications in, e.g., swarm robotics, autonomous vehicles and high-accuracy cancer treatment \cite{wang2012nano, brambilla2013swarm, bechinger2016active}. In fact, the models employed to describe collective behaviours have also been fruitfully exploited in order to build artificial systems with robust behaviors arising from interactions between very simple constituent agents \cite{palacci2013living, rubenstein2014programmable, werfel2014designing}. 

Here, we introduce a novel simple model with short-range aligning interactions between the particles and we study it numerically as a function of the level of orientational noise. First, we study systems constituted by only active particles and we find that there is a transition from a gaseous state at high noise levels to the emergence of metastable clusters at low noise levels. Then, we introduce also passive particles to model the presence of obstacles in the environment and we find a transition towards the emergence of a network of metastable channels along which the active particles can move as the noise level is decreased.

\section{Active Systems}

We consider a system of $N$ active particles moving continuously in a square arena with periodic boundary conditions. The particles are hard spheres with radius $R$ and move with velocity ${\bf v}_n$, where $n=1,...,N$ indicates the particle number. The speed of the particles is assumed constant, i.e. $|{\bf v}_n| \equiv v$. We will perform Brownian dynamics simulations of these particles \cite{volpe2014simulation} where their positions, ${\bf x}_n$, and directions, $\theta_n$, are updated at each time step $t$ according to
\begin{equation} \label{eq:update}
\left\{
\begin{array}{rcl}
{\bf x}_n(t+1) &=& {\bf x}_n(t) + {\bf v}_n(t+1) \\
\theta_n(t+1) &=& \theta_n(t) + T_n + \xi
\end{array}
\right.
\end{equation}
where $\xi$ is a uniformly distributed white-noise term in the interval $[-\eta/2, \eta/2]$ and $T_n$ is a torque term that will be described in the following paragraph. When the volume-exclusion condition is violated so that two particles partially overlap, the particles are separated by moving each one half the overlap distance along their center-to-center axis.

\begin{figure}
\centering
\includegraphics[width=\textwidth]{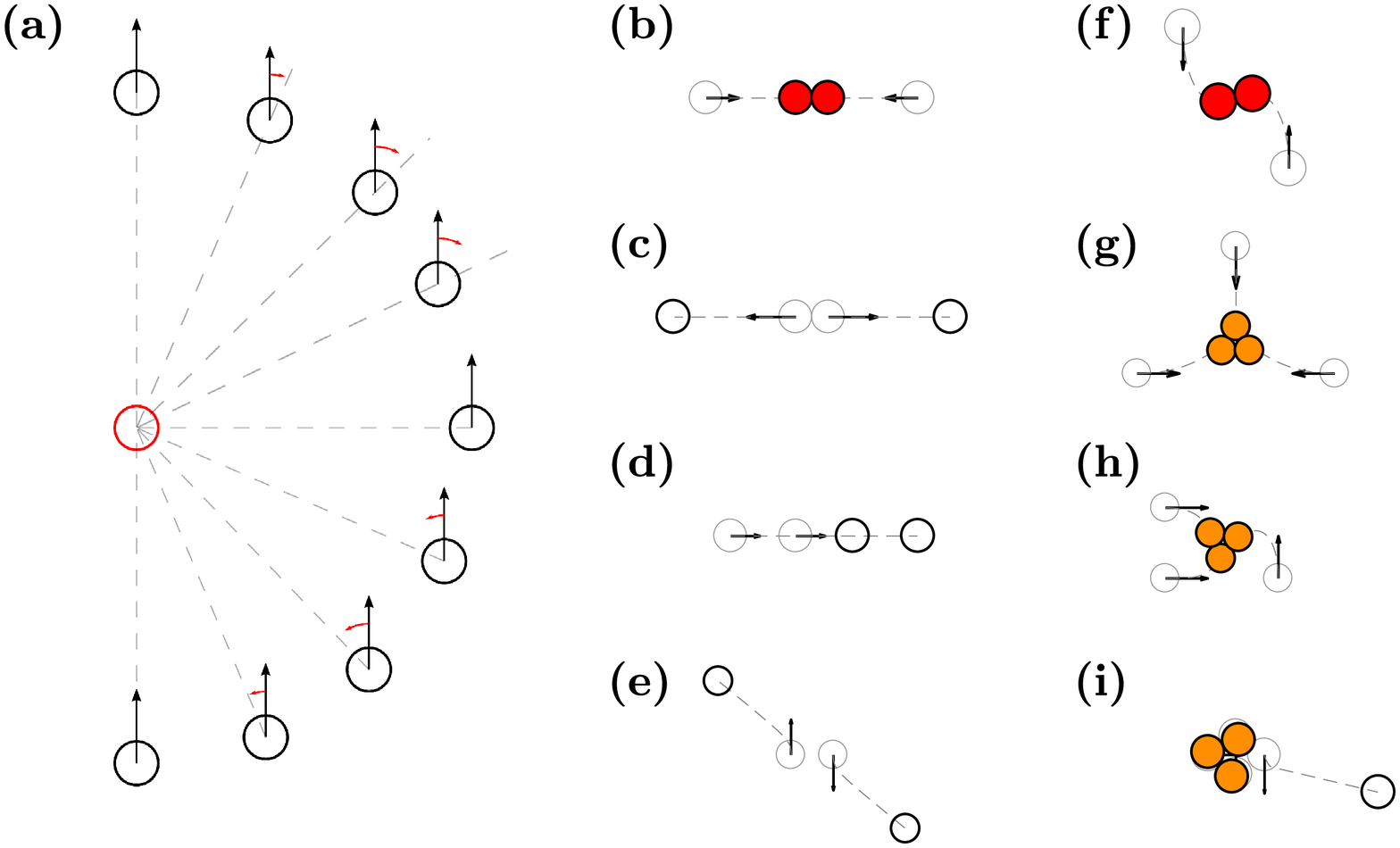}
\caption{{\bf Particle interactions.} 
(a) Torque exerted on the black particle by the red particle: the black arrows denote the direction of motion of the black particle, and the red arrows denote the direction and magnitude of the torque exerted by the red particles.
(b)-(i) Particle behaviours for simple configurations in absence of noise ($\eta=0$). The grey (black) circles denote the initial (final) position of the particles. The arrows represent the initial direction of motion of the particles. 
(b) Two particles oriented towards each other move come together and form a 2-particle cluster. 
(c) Two particles oriented in opposite directions start at contact and move away from each other. 
(d) Two particles oriented in the same direction continue moving in that direction along a straight line. 
(e) Two particles with opposite directions turn until they align and move away from each other along a straight line. 
(f) Two particles turn towards each other until contact and form a 2-particle cluster. 
(g) Three particles initially oriented towards a single point move to form a 3-particle cluster. 
(h) Similar to (g) but with different initial velocities.
(i) In a 4-particle cluster, one particle oriented at an angle of $90^\circ$ away from the centre moves away, while the remaining three particles form a 3-particle cluster.}
\label{fig1}
\end{figure}

The particles exert a torque on each other (Figure~\ref{fig1}a) so that the torque exerted on particle $n$ by all other particles is
\begin{equation} \label{eq:torque}
T_n = T_0 \sum_{i \neq n} \frac{\hat{\bf v}_n \cdot \hat{\bf r}_{ni}}{r_{ni}^2}\; \hat{\bf v}_n \times \hat{\bf r}_{ni} \cdot \hat{\bf e}_z,
\end{equation}
where $T_0$ is a prefactor related to the strength of the interaction, $\hat{\bf v}_n$ is the unit vector representing the direction of motion of particle $n$, $\hat{\bf r}_{ni}$ is the unit vector representing the direction from particle $n$ to particle $i$, $r_{ni}$ is the distance between particle $n$ and  particle $i$, and $\hat{\bf e}_z$ is the unit vector in the direction perpendicular to the plane where the particle moves. Figure~\ref{fig1}a illustrates the effect of this torque: a moving particle (black circle) tends to turn towards another particle (red circle) when the latter is in front, and tends to turn away when the other particle is behind. In the following, we will always set $R=1$, $v=0.05$, and $T_0=1$.

Despite the simplicity of this interaction, it can lead to complex behaviours, including the formation of metastable clusters. Figures~\ref{fig1}b-i show the deterministic motion of particles arranged in simple configurations in the absence of noise ($\eta=0$). Two particles oriented towards each other move along the straight line joining their centres until they come in contact and form a 2-particle cluster (Figure~\ref{fig1}b). If they are oriented away from each other, they move radially away from each other (Figure~\ref{fig1}c). If they are oriented in the same direction, they translate along that direction (Figure~\ref{fig1}d). Two particles with parallel and opposite directions move away from each other (Figure~\ref{fig1}e) or form a 2-particle cluster (Figure~\ref{fig1}f) depending on their initial position. Three particles oriented towards each other form a 3-particle cluster (Figures~\ref{fig1}g-h). In a 4-particle cluster, when one particle gets oriented at an angle of $90^\circ$ away from the centre, it moves away, while the remaining three particles form a 3-particle cluster (Figure~\ref{fig1}i).

\begin{figure}
\centering
\includegraphics[width=\textwidth]{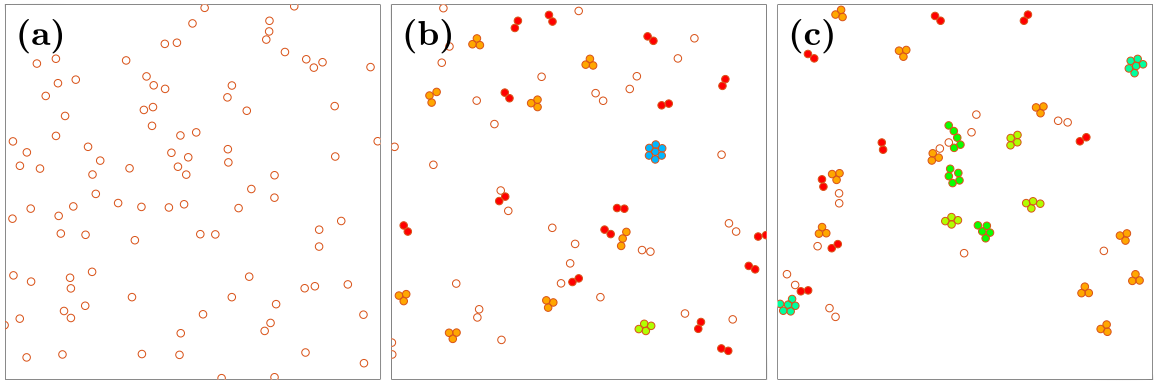}
\caption{{\bf Clustering behaviour.} Steady-state behaviour for a system of $N=100$ active particles for different noise levels: (a) $\eta = 2\pi$, (b) $\eta = 0.2 \pi$ and (c) $\eta = 0.002 \pi$. Particles that belong to a cluster are colour-coded based on the size of the cluster: 2 (red), 3 (orange), 4 (light green), 5 (green), 6 (dark green) and 7 (blue). See also Movie 1 in the supplementary materials.}
\label{fig2}
\end{figure}

Figure~\ref{fig2} shows some snapshots for the steady-state behaviour of the system for different intensities of noise (see also Supplementary Movie 1). For high noise intensity ($\eta=2\pi$, Figure~\ref{fig2}a), the directions of the particles are randomised at each time step and therefore clusters form only rarely, typically comprise only two particles and have a very short lifetime. In this regime the particles interact only when at contact through excluded-volume interactions and are therefore in a gaseous phase. At lower noise levels ($\eta=0.2\pi$, Figure~\ref{fig2}b), persistent clusters start to form, while several free particles are still present. At even lower noise levels ($\eta=0.02\pi$, Figure~\ref{fig2}c), larger clusters form, while almost all particles belong to some of the clusters.

\begin{figure}
\centering
\includegraphics[width=\textwidth]{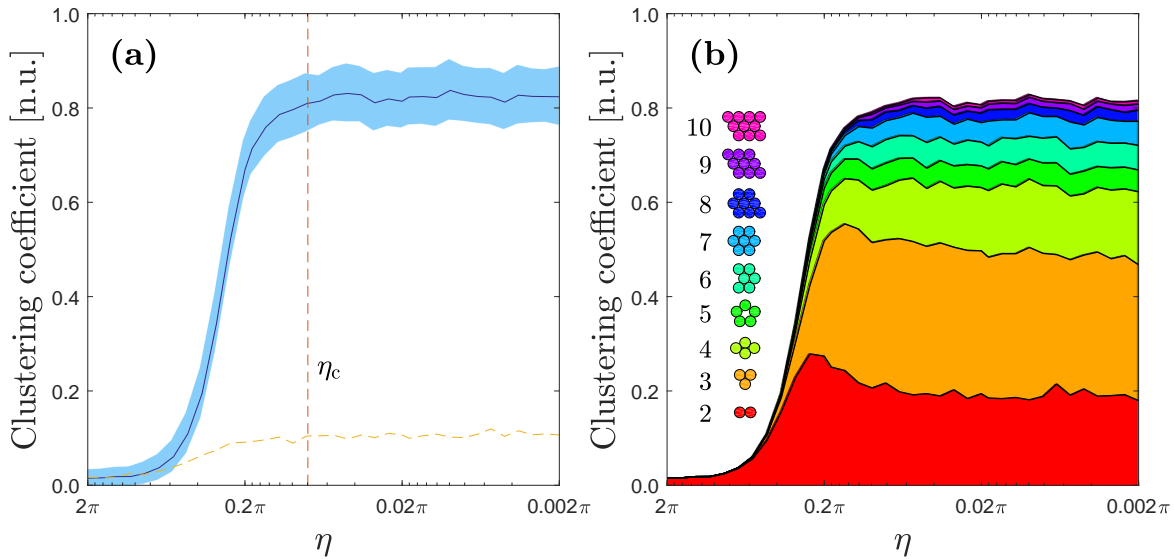}
\caption{{\bf Clustering transition.} 
(a) Clustering coefficient (black solid line) as a function of noise ($\eta$) corresponding to the mean value obtained from 10 simulations of a system of $N=100$ active particles for $10^4$ time steps; the shaded area represents one standard deviation around the mean. The vertical orange dashed line represents the maximum torque exerted on a particle by another at contact $\eta_c = 0.25$ for our parameters (Equation~\eqref{eq:etac}). The purple dashed line represents the average clustering coefficient when $T_0 = 0$.
(b) Cluster size distribution as a function of $\eta$. The colour code corresponds to the sizes of the clusters shown as insets on the right.}
\label{fig3}
\end{figure}

The transition from a gaseous phase to a clustering phase can be seen clearly in Figure~\ref{fig3}a, where the clustering coefficient (defined as the fraction of particles belonging to a cluster) is shown as a function of $\eta$. This transition is related to the balance between the noise level in the system ($\eta$) and the aligning torque between the particles ($T_0$). The maximum noise is $\eta_{\rm max} = \eta/2$. The maximum torque occurs when the particles are at contact and, from Equation~\eqref{eq:torque}, it is $T_{\rm max} = T_0/(8R^2)$. We can expect this transition to occur when $\eta_{\rm max} \approx T_{\rm max}$, which happens at the characteristic noise level
\begin{equation}\label{eq:etac}
\eta_{\rm c} = {T_0 \over 4R^2},
\end{equation}
which is $\eta_{\rm c} = 0.25$ for the parameters of our simulations. This value is represented by the vertical orange dashed line in Figure~\ref{fig3}a and in fact it well-characterises the transition towards the clustered state. The purple dashed line in Figure~\ref{fig3}a represents the average clustering coefficient for the case of particles interacting only through steric interactions ($T_0=0$), which is significanlty below the case of the curve with aligning interactions ($T_0=1$).

Figure~\ref{fig3}b shows the relative abundance of the various cluster sizes as a function of $\eta$. As we have already noted when discussing Figure~\ref{fig2}, as the noise level decreases, at the beginning only small clusters appear, starting with 2-particle clusters, followed by 3-particle clusters and then 4-particle clusters. For higher noise levels, larger clusters become significantly more frequent and, as a consequence, the frequency of smaller clusters decreases.

\begin{figure}
\centering
\includegraphics[width=\textwidth]{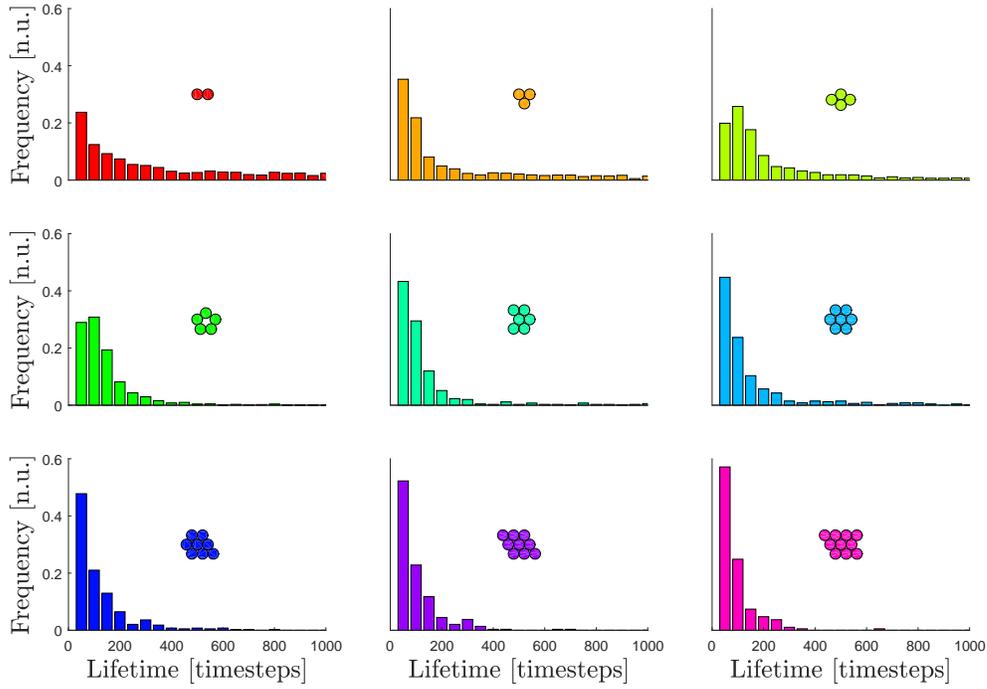}
\caption{{\bf Cluster lifetimes.} 
Distribution of the lifetime of clusters of different sizes (shown as insets) for $\eta = 0.03\pi$.}
\label{fig4}
\end{figure}

\begin{figure}
\centering
\includegraphics[width=0.6\textwidth]{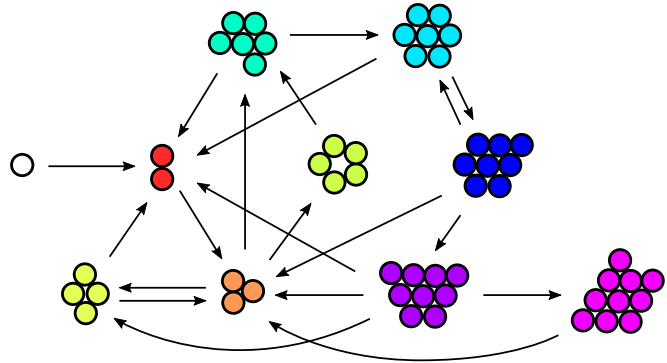}
\caption{{\bf Cluster transitions.} 
The most common transitions between clusters of different sizes are represented by the arrows. The most stable clusters are those of size 2, 3, 6 and 7; in particular, they appear to be stable in the absence of external perturbations, i.e. collisions with other particles. Two-way transitions can be often observed between clusters of sizes 3-4 and 7-8, as shown by the double arrows between them.}
\label{fig5}
\end{figure}

Figures~\ref{fig4} and \ref{fig5} explore the evolution of the clusters as a function of time. Clusters tend to form and grow by collision with single particles and other clusters. These collisions tend to restructure the clusters sometimes leading to a break up into smaller clusters (especially for clusters with more than 6 particles), as can be seen in Supplementary Movie 1.
Figure~\ref{fig4} shows the lifetime of the clusters, which is defined as the number of time steps a cluster exists with a given number of particles until either it loses some of the particles or acquires some extra particles. Smaller clusters (sizes 2, 3 and 4) have typically longer lifetimes, while larger clusters, instead, tend to have a much shorter lifetime. 
Figure~\ref{fig5} shows the most common transitions between clusters. As already observed regarding their lifetimes, the most stable clusters are the ones with 2, 3 and 4 particles. The 2-particle clusters tend to acquire one extra particle and become 3-particle clusters, while only extremely rarely dissolve as two single particles. The 3-particle clusters tend to acquire one extra particle to form 4-particle clusters or to collide with other clusters to form larger clusters (typically, 5-particle or 6-particle clusters). The 4-particle clusters tend to decay by losing one of their particles becoming 3-particle clusters. Larger clusters tend to decay by breaking down into smaller clusters.

\section{Active-Passive Mixed Systems}

It is interesting to explore the mutual influence between active and passive particles in hybrid systems because several natural and artificial systems feature both kinds of particles. In fact, such mixtures have already been explored in several works both experimentally \cite{wu2000particle, koumakis2013targeted, kummel2015formation, pincce2016disorder, argun2016non} and theoretically \cite{gregoire2001active, schwarz2012phase,  stenhammar2015activity}. We will therefore extend the model presented in the previous section by including also $M$ passive particles, which interact with all other (active and passive) particles through volume exclusion and are subject to translational diffusion. We model this diffusion by adding Gaussian white noise to their positions at each time step and set the standard deviation of this noise to $\sigma = 0.1 v$ so that during each time step their displacement is much smaller than that of the active particles. The motion of the active particle is still updated using Eq.~\eqref{eq:update}, with the only difference that now the torque is given by 
\begin{equation}
T_n = T_0 \sum_{i \neq n} \frac{\hat{\bf v}_n \cdot \hat{\bf r}_{ni}}{r_{ni}^2} \; \hat{\bf v}_n \times \hat{\bf r}_{ni} \cdot \hat{\bf e}_z - T_0 \sum_{m} \frac{\hat{\bf v}_n \cdot \hat{\bf r}_{nm}}{r_{nm}^2} \; \hat{\bf v}_n \times \hat{\bf r}_{nm} \cdot \hat{\bf e}_z ,
\end{equation}
where $m = 1,...,M$ is the index of the passive particles and ${\bf r}_{nm}$ is the relative position vector between particle $n$ and obstacle $m$. The sign of the torque due to the passive particles is negative so that the active particles tend to avoid the passive ones. This kind of behaviour is observed, for example, in the motion of people across a stationary crowd \cite{helbing2005self} and in the motion of microswimmers in the presence of obstacles \cite{takagi2014hydrodynamic, spagnolie2015geometric}.

At low concentrations of passive particles ($M < N$) and at low packing fractions, the overall qualitative behaviour of the active particles is unaffected, so that the active particles still cluster as described in the previous section.

\begin{figure}
\centering	
\includegraphics[width=\textwidth]{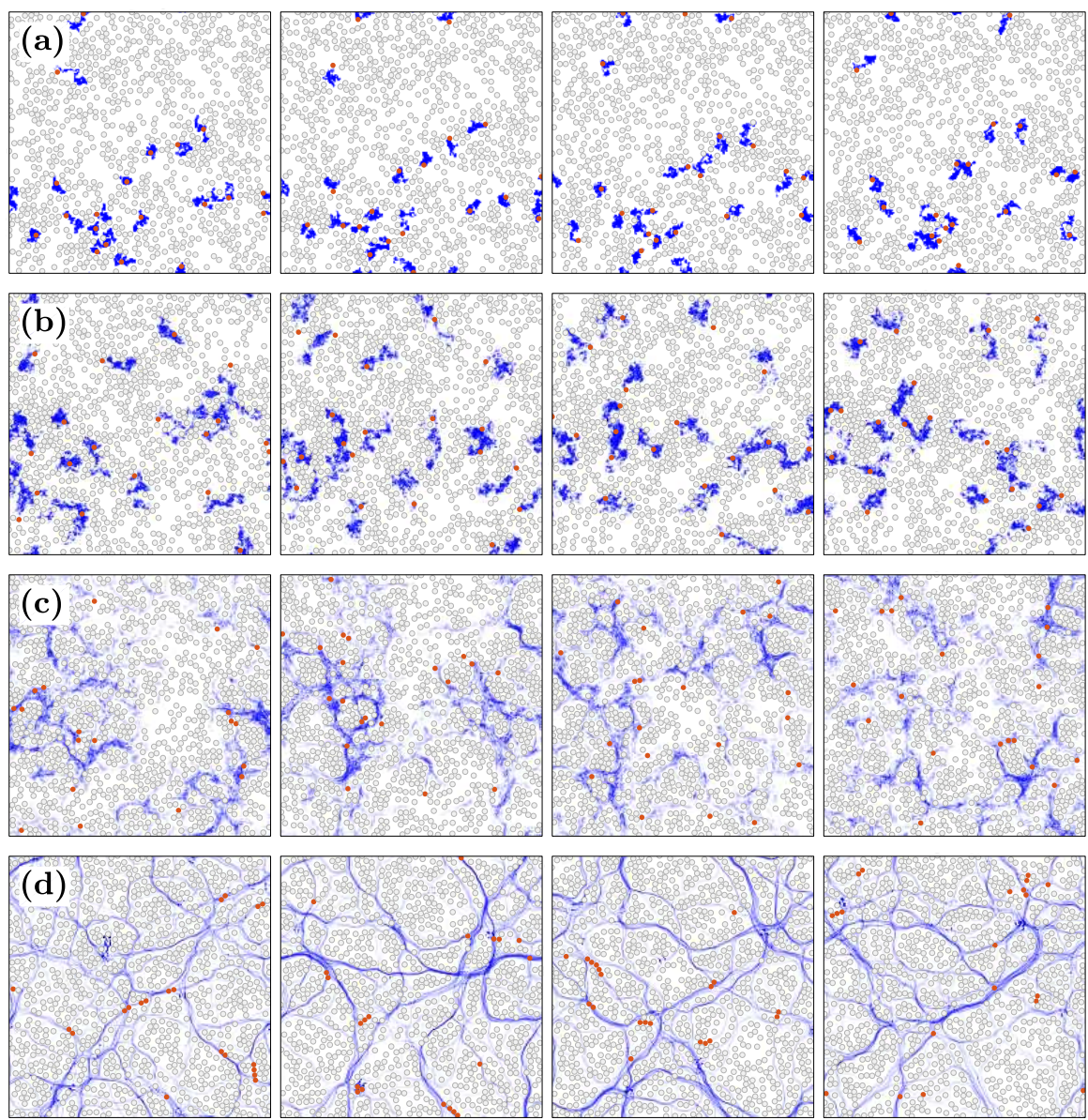}
\caption{{\bf Metastable channel formation.} 
Simulation of a system of 20 active particles (red circles) and 900 passive particles (grey circles) for different levels of noise: (a) $\eta = 2\pi$, (b) $= \pi$, (c) $= 0.5 \pi$ and (d) $\eta = 0.03\pi$. From left to right, the plots correspond to time steps $t = 25\,000$, $50\,000$, $75\,000$ and $100\,000$. The blue shades represent the trails left by the passage of the active particles over the preceding $25\,000$ time steps. In (d), thanks to the low noise level, metastable channels are opened and are stabilized by the active particles. See also Movie 2 in the supplementary materials.}
\label{fig6}
\end{figure}

\begin{figure}
\centering
\includegraphics[width=0.5\textwidth]{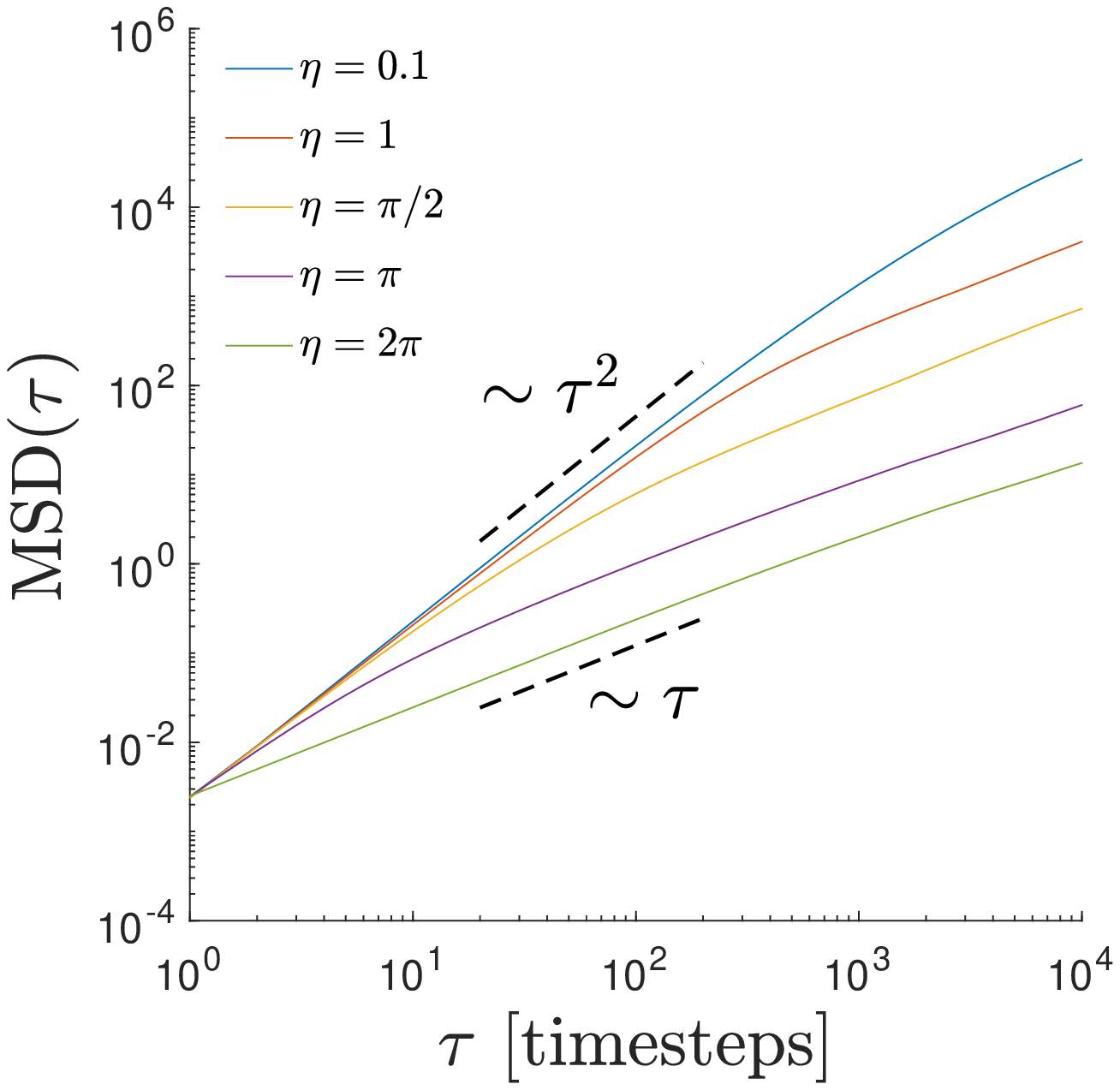}
\caption{{\bf Mean square displacement (MSD) of active particle in the presence of passive particles.} 
MSD of active particles in the presence of a background of passive particles in the conditions shown in Figure~\ref{fig6} as a function of the noise level ($\eta$). The MSDs feature a transitions from ballistic motion at small times ($\mathrm{MSD}(\tau) \propto \tau^2$ for small $\tau$) to diffusive motion at long times ($\mathrm{MSD}(\tau) \propto \tau$ for large $\tau$). The crossover time between the two regimes increases as the noise level decreases.}
\label{fig7}
\end{figure}

More interesting phenomena occur at high packing fractions and high numbers of passive particles compared to the active ones ($M \gg N$). The behaviour of one such system is shown in Figure~\ref{fig6} and Supplementary Movie 2 for various noise levels. At high noise levels ($\eta=2\pi$, Figure~\ref{fig6}a) the motion of the active particles is significantly hindered by the presence of the passive particles and is essentially diffusive, as shown by the mean square displacement (MSD) which has a slope of 1 at all times (green line in Figure~\ref{fig7}). The active particles compress the passive ones creating some voids within the background of passive particles, within which they are effectively confined. As a consequence, the active particles have few chances of encountering each other and forming clusters. This behaviour is similar to that observed in experiments with microswimmers in a bath of passive particles, where it was observed that even the presence of very few active particles could help the crystallisation of the system \cite{kummel2015formation}.

Decreasing the noise level to $\eta=\pi$ (Figure~\ref{fig6}b), the active particles are able to move more, but remain confined within depleted regions created in the background of passive particles. Even though they can perform some straight runs, they quickly get blocked by the passive particles and their movement becomes quickly diffusive. This is reflected by the MSD shown by the purple line in Figure~\ref{fig7}, which after a brief superdiffusive stage for short times (${\rm MSD}(\tau) \propto \tau^2$ for small $\tau$)  quickly become diffusive (${\rm MSD}(\tau) \propto \tau$ for large $\tau$). Also in this case, the passive particles present in the background prevent encounters between active particles and the formation of clusters. 

A similar behaviour takes place even at even lower noise levels ($\eta=0.5\pi$, Figure~\ref{fig6}c). Again, the motion is directed only for short time scales and quickly becomes diffusive (see corresponding MSD plotted by the yellow line in Figure~\ref{fig7}). However, one can now observe the formation of some proto-channels, which are highlighted by the shaded blue areas which represent the trails left by the passage of the active particles over $25\,000$ time steps preceding the one represented in the figure. These proto-channels permit the active particles to occasionally encounter each other and form some 2-particle clusters.

Decreasing further the noise level to $\eta = 0.03 \pi$ (Figure~\ref{fig6}d) leads to the formation of fully-fledged channels, whose presence is clearly shown by the blue shaded areas. These are open areas free of passive particles where the active particles can propagate unhindered. Comparing the various panels in Figure~\ref{fig6}d, which correspond to different times and represents trails corresponding to non-overlapping time frames, one can see that these channels are quite stable over time. The reason for this is that, once a channel is open by some active particles, additional  active particle use it leading to its dynamic stabilisation. This transition towards the formation of channels is driven by the increase of the characteristic length of the directed runs of the active particles in the background of passive particles. As can be seen from the blue line in Figure~\ref{fig7}, the MSD is now ballistic over a longer time range. Once the channels are open the particles can encounter each other and form some small (2-particle and 3-particle) clusters.

\section{Conclusions and Outlook}

We have introduced a novel simple model for interacting active particles that leads to the formation of metastable clusters and, in the presence of a background of passive particles, to the opening of metastable channels. This model can be easily implemented in artificial systems based on robots equipped with sensors thanks to its simplicity and the fact that it relies only on the knowledge of the positions of the surrounding particles. Furthermore, by changing the level of noise in real time, it would be possible to switch between different behaviours, e.g. between clustering and dispersion.

\section{Acknowledgements}

This work was partially supported by the ERC Starting Grant ComplexSwimmers (grant number 677511).

\end{document}